\documentclass[aoas,nameyear,seceqn,dvips]{arximspdf}
\usepackage{dcolumn,graphics}

\doi{10.1214/10-AOAS336}
\volume{4}
\issue{3}
\pubyear{2010}
\firstpage{1533}
\lastpage{1557}

\makeatletter
\newcolumntype{d}[1]{D{.}{.}{#1}}

\newtheorem{prop}{Proposition}
\newproclaim{example}{Example}
\makeatother

\begin{document}
\begin{frontmatter}

\title{Bayesian inference for double Pareto\\ lognormal queues}
\runtitle{Inference for double Pareto lognormal queues}

\begin{aug}
\author[a]{\fnms{Pepa} \snm{Ramirez-Cobo}\corref{}\thanksref{t1}\ead[label=e1]{Pepa.Ramirezcobo@gipsa-lab.inpg.fr}},
\author[b]{\fnms{Rosa E.} \snm{Lillo}\ead[label=e2]{lillo@est-econ.uc3m.es}},\\
\author[c]{\fnms{Simon} \snm{Wilson}\ead[label=e3]{Simon.Wilson@tcd.ie}}
\and
\author[b]{\fnms{Michael P.} \snm{Wiper}\ead[label=e4]{mwiper@est-econ.uc3m.es}}

\thankstext{t1}{Supported in part by the Comunidad de Madrid-UC3M
(Project 2006/03565/001).}

\runauthor{Ramirez-Cobo, Lillo, Wilson and Wiper}

\affiliation{CNRS France, Universidad Carlos III de Madrid, Trinity
College Dublin and Universidad Carlos III de Madrid}

\address[a]{P. Ram\'irez-Cobo\\
Laboratoire Images et Signaux\\
\quad(UMR CNRS 5083)\\
ENSEIG, BP 46, 38402\\
Saint-Martin d'H\`eres\\
France\\
\printead{e1}}

\address[b]{R. E. Lillo\\
M. P. Wiper\\
Departamento de Estad\'istica\\
Universidad Carlos III de Madrid\\
Spain\\
\printead{e2}\\
\phantom{E-mail:} \printead*{e4}}

\address[c]{S. Wilson\\
Department of Statistics\\
Trinitiy College Dublin\\
Ireland\\
\printead{e3}}
\end{aug}

\received{\smonth{7} \syear{2009}}
\revised{\smonth{1} \syear{2010}}

%
\begin{abstract}
In this article we describe a method for carrying out Bayesian
{estimation} for the double Pareto lognormal (\textit{dPlN})
distribution which has been proposed as a model for
heavy-tailed phenomena. We apply our approach to estimate the
$\mathit{dPlN}/M/1$ and $M/\mathit{dPlN}/1$ queueing systems. These
systems cannot
be analyzed using standard techniques due to the fact that the
\textit{dPlN} distribution does not possess a Laplace transform in
closed form. This difficulty is overcome using some recent
approximations for the Laplace transform of the interarrival
distribution for the $\mathit{Pareto}/M/1$ system. Our procedure is illustrated
with applications in internet traffic analysis and risk theory.
\end{abstract}

%

\end{frontmatter}
%

\section{Introduction}
\label{sec-Introduction}
Heavy-tailed distributions have been used to model a variety of
phenomena in areas such as economics, finance, physical and
biological problems; see Adler, Feldman and Taqqu (\citeyear{Adler}). In
particular, a number
of variables in teletraffic engineering, such as file sizes, packet
arrivals, etc., have been shown to possess heavy-tailed
distributions; {this can be found, for example, in} Paxson and Floyd
(\citeyear{Paxson}). Also, in an actuarial context, insurance claim sizes can often
be very large and in such
cases, may be modeled as long tailed; see, for example, Embrechts, Kl\"
{u}ppelberg and Mikosch (\citeyear{Embrechts}). For a detailed review of heavy-tailed
distributions, {we refer the reader to} Sigman (\citeyear{Sigman}).

The Pareto distribution has often been applied to model the heavy-tail
behavior of teletraffic variables [Resnick (\citeyear{Resnick})] and insurance
claims [Philbrick (\citeyear{Philbrick})]. In particular, in Ramirez, Lillo and Wiper
(\citeyear{Ramirez}) a mixture of $k$ Pareto distributions ($k$-$\mathit{Par}$) is used to
model ethernet packets interarrival times. However, although the
Pareto distribution often models the tails of a distribution well, it
is unimodal and decreasing, which means that it will not model the body
of the distribution correctly in many modeling situations as is shown
in some of the examples in this paper.

Reed and Jorgensen (\citeyear{Reed}) recently introduced the double Pareto
lognormal (\textit{dPlN}) distribution as a versatile model for heavy-tailed
data and considered various frequentist approaches to inference for
this distribution. They did not recommend the method of moments as an
estimation method, and observed that the EM algorithm sometimes
encounters convergence problems. In this work we focus on the Bayesian
approach, which may be preferred for problems where the interest is not
only in inference but also in prediction; see, for example, Robert
(\citeyear{Robert}). The first
objective of this paper is thus to develop an algorithm to implement
Bayesian inference for the \textit{dPlN} distribution.

The study of congestion in teletraffic systems and of ruin problems in
insurance is directly related to the analysis of queueing
systems, where the arrival or service process are defined by a
heavy-tailed distribution. In this paper we consider the $\mathit{dPlN}/M/1$ and
$M/\mathit{dPlN}/1$ queues, which, to our knowledge, have not been considered
before in the literature.

The usual moment generating function approach to obtaining the
equilibrium distribution of a queue [Gross and Harris (\citeyear{Gross})] is
difficult to implement because the \textit{dPlN} distribution lacks a
moment generating function in closed form. An alternative, which we
shall apply, is based on a direct
approximation of the nonanalytical Laplace transform using a
variant of the transform approximation method (\textit{TAM}); see Harris
and Marchal (\citeyear{Harris98}), \citet{Harris00} and Shortle et al. (\citeyear{Shortle04}). The
first version of the \textit{TAM}, known as \textit{Uniform TAM} or \mbox
{\textit{U-TAM}}, was implemented in \citet{Ramirez}, where
estimation of the \mbox{$k$-$\mathit{Par/M/\mathrm{1}}$} queue was considered. In this
paper we propose a variant of the \textit{TAM} based on both the \emph
{Uniform} and \textit{Geometrical} \textit{TAM}s. By
combining this variant of the \textit{TAM} with the Bayesian inference
method for the \textit{dPlN}
distribution, we can obtain {estimates} of queueing
properties of interest such as the probability of congestion.

This paper is organized as follows. In Section \ref{sec2} we review the
definition and key properties of the \textit{dPlN} distribution and
present an approach to Bayesian inference for this distribution,
illustrating our procedure with simulated and real data. In Section \ref{sec3} we
examine the $\mathit{dPlN}/M/1$ queueing system and show how the \textit{TAM} approach
can be used to approximate the Laplace transform of the
\textit{dPlN} distribution. Our results are then applied to a real
example of internet traffic arrivals. In Section \ref{sec4} we study the
$M/\mathit{dPlN}/1$ queueing system and show how the waiting time distribution
of this system can be estimated. We then apply our results to the
estimation of the ruin probability given real insurance claims data.
Conclusions and possible extensions to this work are considered in
Section \ref{sec5}.
%
\section{Bayesian inference for the double Pareto lognormal
distribution}\label{sec2}

\subsection{The double Pareto lognormal distribution}
A random variable $Y$ is said to have
a \textit{Normal Laplace distribution} (\textit{NL}), denoted $Y\sim
\mathit{NL}(\alpha,\beta,\nu,\tau^2)$ if $Y = Z + W$, where
$Z\sim
N(\nu,\tau^2)$,
and $W$ is a skewed Laplace distributed variable with density function
\[
f_{W}(w|\alpha,\beta)=\cases{\displaystyle
\frac{\alpha\beta}{\alpha+\beta}e^{\beta w}, & \quad $\mbox{if } w\leq
0,$ \cr
\displaystyle\frac{\alpha\beta}{\alpha+\beta}e^{-\alpha w}, & \quad $\mbox{if }w>0,$}
\]
independent of $Z$, for $\alpha,\beta>0$. The density
function of $Y$ is
\begin{eqnarray*}
f_Y(y|\alpha,\beta,\nu,\tau^2)&=& \frac{\alpha\beta
}{\alpha
+\beta}\phi\biggl(
\frac{y-\nu}{\tau} \biggr)\\
&&{}\times \bigl[ R\bigl(\alpha\tau-(y-\nu)/\tau\bigr) +
R\bigl(\beta\tau+ (y-\nu)/\tau\bigr) \bigr], \label{pdfY}
\end{eqnarray*}
where $R(z)$ is the Mill's ratio defined by
%
%
\begin{equation}\label{mills}
R(z)=\Phi^c(z) /\phi(z),
\end{equation}
where
$\Phi^c(z)=1-\Phi(z)$ and $\phi(z)$ and $\Phi(z)$ are the standard
normal density and cumulative distributions respectively.

A random variable, $X$, is said to have a \textit{dPlN} distribution
with parameters
$(\alpha,\beta,\nu,\tau^2)$ if $X=\exp(Y)$ where $Y$ is
Normal Laplace distributed.

The usual change of variable to the density of $Y$ gives the density of
$X$ to be
\begin{eqnarray*}\label{pdfX}
f_X(x|\alpha,\beta,\nu,\tau^2 )&=& \frac{\alpha\beta
}{\alpha
+\beta} \biggl(\frac{1}{x}\biggr) \phi\biggl(
\frac{\log x-\nu}{\tau} \biggr) \\
&&{}\times\bigl[ R\bigl(\alpha\tau-(\log
x-\nu)/\tau\bigr) + R\bigl(\beta\tau+ (\log x-\nu)/\tau\bigr) \bigr].
\end{eqnarray*}

Also, Reed and Jorgensen (\citeyear{Reed}) show that the $\mathit{dPlN}(\alpha,\beta,\nu
,\tau^{2})$ can be represented as a mixture as
\[
f_X(x|\alpha,\beta,\nu,\tau^2 )=\frac{\beta}{\alpha
+\beta
}f_{1}(x|\alpha,\nu,\tau^2 )+\frac{\alpha}{\alpha
+\beta
}f_{2}(x|\beta,\nu,\tau^2 ),
\]
where the densities
%
%
\begin{eqnarray}\label{f1}
f_{1}(x|\alpha,\nu,\tau^2 )&=& \alpha x^{-\alpha-1}
e^{\alpha\nu+\alpha^{2}\tau^2/2}\ \Phi\biggl(\frac{\log(x)-\nu
-\alpha
\tau^{2}}{\tau}\biggr),\\ \label{f2}
f_{2}(x|\beta,\nu,\tau^2 )&=&\beta x^{\beta
-1}e^{-\beta\nu
+\beta^{2}\tau^2/2}\ \Phi\biggl(\frac{\log(x)-\nu+\beta\tau
^{2}}{\tau
}\biggr)
\end{eqnarray}
are, respectively, the limiting forms (as $\beta\rightarrow\infty$
and $\alpha\rightarrow\infty$) of the $\mathit{dPlN}(\alpha,\beta,\break\nu,\tau
^{2})$ distribution.

Reed and Jorgensen (\citeyear{Reed}) illustrate the form of the \textit{dPlN}
density function for various different groups of parameter values. In
particular, they show that it exhibits upper
power-tail behavior in that $f_X(x) \rightarrow k x^{-\alpha-1}$
as $x \rightarrow\infty$. The \textit{dPlN} distribution does not
possess a moment generating function in closed form. However, if
$r<\alpha$, the moment of order $r$ exists:
\[
E(X^r\vert\alpha, \beta, \nu, \tau^2)=\frac{\alpha
\beta
}{(\alpha-r)(\beta+r)}e^{r\nu+ r^2 \tau^2/2}.
\]
Reed and Jorgensen (\citeyear{Reed}) also illustrate a procedure for frequentist
inference for
the \textit{dPlN} distribution using the EM algorithm and note
that under certain conditions, this approach suffers from problems
of convergence. An alternative procedure which has not been
examined thus far is to take a Bayesian approach, as we do here.

\subsection{Bayesian inference}
Given a random sample $\mathbf{x}=(x_1,\ldots,x_n)$ from the \textit
{dPlN}($\alpha, \beta, \nu, \tau^2$), the
goal is to compute a posterior distribution\break \mbox{$f(\alpha, \beta,
\nu, \tau^2  | \mathbf
{x})$}. For ease of notation, we define $\bolds{\theta}=
(\alpha, \beta, \nu, \tau^2)$ in what follows. It is easier
computationally to work with
the normal Laplace, hence, we define $\mathbf{y}=(y_1,\ldots,y_n)$,
where $y_i=\log(x_i)$, $i=1,\ldots,n$, and
compute the posterior density function $f(\bolds{\theta}  |
\mathbf{y})$
using the normal Laplace likelihood.

The definition of a normal Laplace random variable $Y\sim \mathit{NL}
(\alpha
,\beta,\nu,\tau^2)$ suggests
the use of a Gibbs sampler where one considers the two components of $Y$
as auxiliary variables to be sampled along with $\bolds{\theta}$ so that
sampling $\bolds{\theta}$ then reduces to sampling $(\alpha,\beta)$ and
$(\nu,\tau^2)$ from distributions with truncated skewed
Laplace and
Gaussian likelihoods respectively. The classical EM algorithm
developed in Reed and Jorgensen (\citeyear{Reed}) was based on a similar idea,
but, as noted earlier, this can show convergence problems.

The conditional distribution of $Z|Y=y,\alpha, \beta, \nu, \tau^2$
is a mixture of two truncated normal variables
as stated in the following proposition.
\begin{prop}\label{prop1}
The conditional distribution of $Z|Y,\alpha, \beta, \nu, \tau^2$
is\break
a weighted mixture of
two truncated normal densities:
%
%
\begin{eqnarray}\label{conddens}
\qquad f_{Z|y}(z|y,\alpha, \beta, \nu, \tau^2)&=&
\biggl(R(y_{\beta})\frac{\phi(z^\beta
)}{\tau\Phi^c(y^\beta)} I_{z\ge y} + R(y_{\alpha})\frac{\phi
(z^\alpha
)}{\tau\Phi^c(y^\alpha)}I_{z<y} \biggr)
\nonumber
\\[-8pt]
\\[-8pt]
\nonumber
&&{}\Big/\bigl(R(y_{\alpha})+R(y_{\beta})\bigr),
\qquad
z \in\mathbb{R},
\end{eqnarray}
where $R(\cdot)$ is given in \textup{(\ref{mills})}, and
\begin{eqnarray*}
y_{\alpha}&=& \alpha\tau- (y-\nu)/\tau, \qquad y_{\beta}=\beta\tau+
(y-\nu)/\tau,
\\
y^\alpha&=&\frac{y-(\nu+\tau^2 \alpha)}{\tau},\qquad  y^\beta= \frac
{y-(\nu-\tau^2 \beta)}{\tau},
\\
z^\alpha&=& \frac{z-(\nu+\tau^2 \alpha)}{\tau}, \qquad  z^\beta=
\frac
{z-(\nu-\tau^2 \beta)}{\tau}.
\end{eqnarray*}
\end{prop}

For a proof of Proposition \ref{prop1} see Appendix \hyperref[appa]{A}.

Note now that we can express the skewed Laplace
distribution as the difference of two exponential variables, that
is,
\[
W = E_1-E_2 \qquad \mbox{where } E_1| \alpha\sim\mathcal{E}(\alpha) \mbox{ and }
E_2|\beta\sim\mathcal{E}(\beta).
\]

The following proposition specifies the conditional distribution of $E_1|W$.
\begin{prop}\label{prop2}
The distribution of $E_1|W,\alpha,\beta$ is a truncated exponential
with support $[\max\{w,0\},\infty)$,
%
%
\begin{equation}\label{conddens2}
f_{E_1|W}(e_1|w,\alpha,\beta)=\frac{(\alpha+\beta)e^{-(\alpha
+\beta
)e_1}}{I_{w<0}+e^{-(\alpha+\beta)w}I_{w\ge
0}}
\end{equation}
for $e_1 > \max\{w,0\}$.
\end{prop}

The proof of Proposition \ref{prop2} can be found in Appendix \hyperref[appb]{B}.
Given a sample, $(y_1,\ldots,y_n)$
conditional on the parameters $(\alpha,\beta, \nu, \tau
^2),$
then we
can generate $(z_1,\ldots, z_n)$ from the formula in Equation
(\ref{conddens}). Also, we can define $\mathbf{ w}=\mathbf{ y}-\mathbf{
z}$, $w_1=y_1-z_1$, $\ldots,$ $w_n=y_n-z_n$ and then generate
$\mathbf{e}_1=(e_{1,1},\ldots,e_{1,n})$ from the formula in Equation
(\ref
{conddens2}) and
define $\mathbf{e}_2 = \mathbf{e}_1-\mathbf{ w}$. To undertake inference for $\nu
$, and $\tau^2$, let us suppose that we use a normal, inverse gamma
prior distribution
%
%
\begin{eqnarray}\label{priornu}
\nu|\tau^2 & \sim& \mathcal{N}\biggl(m,\frac{\tau^2}{k}\biggr), \\
\label{priortau}
\frac{1}{\tau^2} & \sim& \mathcal{G}\biggl(\frac{a}{2},\frac{b}{2}\biggr).
\end{eqnarray}
Then, from standard Bayesian theory [see, e.g., Box and Tiao (\citeyear{boxtiao})],
\begin{eqnarray*} \nu\vert\tau^2,\mathbf{ z} & \sim& \mathcal{N}
\biggl(\frac{km+n\bar{z}}{k+n},
\frac{\tau^2}{k+n} \biggr), \\
\frac{1}{\tau^2}\big|\mathbf{z} & \sim
& \mathcal{G}\biggl(\frac{a+n}{2},\frac{b+(n-1)s^2_z +
(kn/(k+n))(m-\bar{z})^2}{2} \biggr),
\end{eqnarray*}
where $\bar{z}=\sum_{i=1}^n z_{i}/n$ and $s^2_z=\sum_{i=1}^n \sum
(z_{i}-\bar{z})^2/(n-1)$. Also, given gamma priors $\alpha\sim\mathcal{G}(c_{\alpha},d_{\alpha})$, $\beta\sim\mathcal{G}(c_{\beta},d_{\beta})$, then
%
\begin{eqnarray}\label{prioralpha}
\alpha|\mathbf{e}_1 & \sim& \mathcal{G}(c_{\alpha}+n,d_{\alpha}+n
\bar{e_1}), \\ \label{priorbeta}
\beta|\mathbf{e}_2 & \sim& \mathcal{G}(c_{\beta}+n,d_{\beta}+n \bar{e_2}).
\end{eqnarray}
{Of course, many other prior structures are possible. In particular, it
might be assumed that $\nu$ and $\tau^2$ are independent a priori, or
that there is some prior dependence between $\alpha,\beta$ and $\nu
,\tau
^2$. In the presence of real prior information, the use of such
alternative structures could lead to more flexible modeling. However,
the main disadvantage is that the semi conjugate structure implied
given the proposed prior distributions is lost and more complex MCMC
algorithms would have to be used to undertake inference.}

{Therefore, we can define the following Gibbs algorithm}: \texttt{
\begin{enumerate}
\item[1.] Set initial values
$\alpha^{(0)},\beta^{(0)},\nu^{(0)},{\tau^2}^{(0)}$.
\item[2.] For
$t=1,\ldots,T$
\begin{enumerate}
\item[a.] For $i=1,\ldots,n$,
\begin{enumerate}
\item[a1.] Generate $z_i^{(t)}$ from
$f(z|\alpha^{(t-1)},\beta^{(t-1)},\nu^{(t-1)},{\tau
^2}^{(t-1)}\vert
y_i)$.
\item[a2.] Set $w_i^{(t)}=y_i-z_i^{(t)}$.
\item[a3.]
Generate $e_{1,i}^{(t)}$ from
$f(e_1|w_i,\alpha^{(t-1)},\beta^{(t-1)})$.
\item[a4.]
Set $e_{2,i}^{(t)} = e_{1,i}^{(t)}-w_i^{(t)}$.
\end{enumerate}
\item[b.] Generate ${\tau^2}^{(t)} \sim f(\tau^2|\mathbf{
z}^{(t)})$.
\item[c.] Generate $\nu^{(t)} \sim
f(\nu|\mathbf{ z}^{(t)},{\tau^2}^{(t)})$.
\item[d.] Generate
$\alpha^{(t)} \sim f(\alpha|\mathbf{e}_1^{(t)})$.\vspace*{1pt}
\item[e.] Generate $\beta^{(t)} \sim f(\beta|\mathbf{e}_2^{(t)})$.
\end{enumerate}
\end{enumerate}
}

In the presence of little prior information, it would appear natural
to use a noninformative, improper prior distribution. However, it is
easy to show that
in this case, the posterior distribution is also improper.
\begin{prop}\label{prop3} If an improper prior distribution for
$\alpha
$ and $\beta$ is used in the sense that $\int_{a}^{\infty} f(\alpha
|\beta)  \, d\alpha$ diverges for all $a \ge0$, $\beta>0$ or $\int
_{b}^{\infty} f(\beta|\alpha) \,  d\beta$ is a divergent integral for
any $ b\ge0$, $\alpha>0$, then the posterior distribution is also improper.
\end{prop}

The proof of Proposition \ref{prop3} can be found in Appendix \hyperref[appc]{C}. This implies
that in order to carry out Bayesian inference, it is fundamental to use
a proper prior distribution for $\alpha,\beta$.

\subsection{Illustration with simulated and real data sets}\label{subsec:illustration}

\begin{example}\label{ex1}
As an illustration of the proposed Gibbs sampler with simulated data,
consider a sample of size 1000, generated from $\mathit{dPlN}(0.25,0.5,1,1)$. The
Gibbs algorithm was run for 500,000 iterations with initial
values set to $\bolds{\theta}^{(0)}=(0.2625, 0.5529, 1.1992, 0.8147)$, the maximum
likelihood estimates. {The hyperparameters were set to $m=0$, $k=4$ in
(\ref{priornu}), $a=b=1$ in (\ref{priortau}) and $c_{\alpha
}=c_{\beta
}=d_{\alpha}=d_{\beta}=1$ in (\ref{prioralpha})--(\ref{priorbeta}), and
from now on these are the values used in the rest of the examples}. In
order to avoid high autocorrelation, we
did thinning and took one sample out of 50. Gibbs sampler code was
written in Matlab and, when run on Intel Core Duo at 2.4 GHz and 2 GB of
DDR3 RAM, took approximately 19 minutes to perform 100,000 iterations.
Figure \ref{Fig4.1} illustrates the mixing properties of the
algorithm.
%
\begin{figure}

\includegraphics{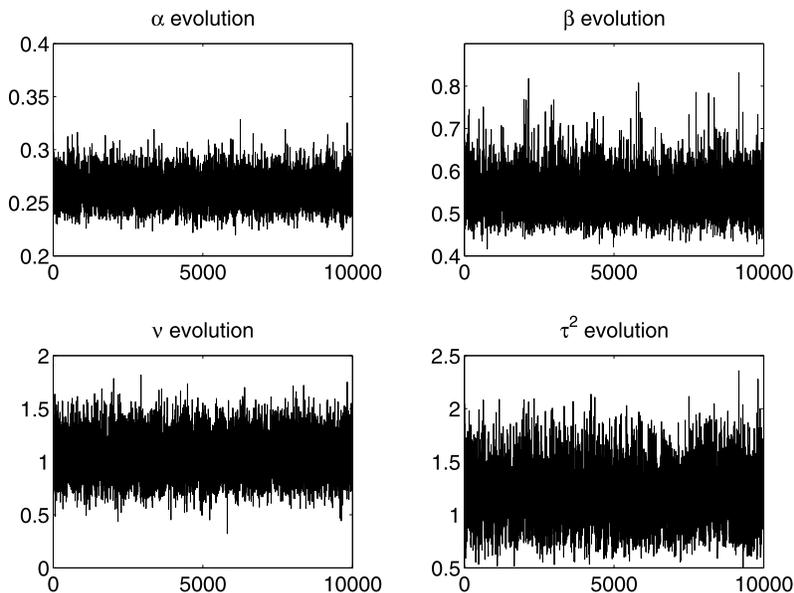}

\caption{MCMC trace plots for Example \protect\ref{ex1}. The Gibbs
algorithm was applied to a sample of 1000 data generated from a $\mathit{dPlN}$
distribution with parameters $\bolds{\theta}=(0.25,0.5,1,1)$.}\label{Fig4.1}
\end{figure}
We found $E(\bolds{\theta}|\mathbf{ y})=(0.2578,0.4995,1.065,1.1848)$
close to
the maximum
likelihood estimates. In addition, we computed credible intervals and
correlations in the posterior as measures of precision of the
estimates. Credible intervals ($95\%$) for the parameters $\alpha$
$\beta$, $\nu$ and $\tau^2$ were
\begin{eqnarray*}
C_{\alpha}&=&[0.2377, 0.2906],\qquad   C_{\beta}=[0.4702, 0.6401],\\
C_{\nu}&=&[0.7044, 1.4409], \qquad  C_{\tau^2}=[0.7178, 1.7352].
\end{eqnarray*}
With respect to the posterior correlations, we found
\[
\pmatrix{
& \alpha& \beta& \nu& \tau^2 \cr
\alpha& 1 &\ 0.1449 &\ 0.5525 &\ 0.5568 \cr
\beta& &\ 1 &\ -0.4936 &\ 0.4534 \cr
\nu& & &\ 1&\ 0.2362 \cr
\tau^2 & & & &\ 1 \cr
}
.
\]
%
%
Notice that, for example, the parameters $\beta$ and $\nu$ are
negatively correlated a posteriori and $\alpha$ and $\nu$ positively, a
consequence of the definition of a Normal-Laplace distribution as the
sum of a normal and skewed Laplace variables.

In Figure \ref{Fig4.4} the {fitted density function, estimated for the
data (in log-scale), and almost undistinguishable from the theoretical
one, is depicted. }{ The fitted curve has been computed by simple
averaging over the Gibbs sampled values, that is, ${f}_{Y}(y|\mathbf{ y})$
has been estimated by
\[
\frac{1}{T} \sum_{t=1}^T f_{Y}\bigl(y|\alpha^{(t)},\beta^{(t)},\nu
^{(t)},{\tau^2}^{(t)}\bigr).
\]
}
\begin{figure}

\includegraphics{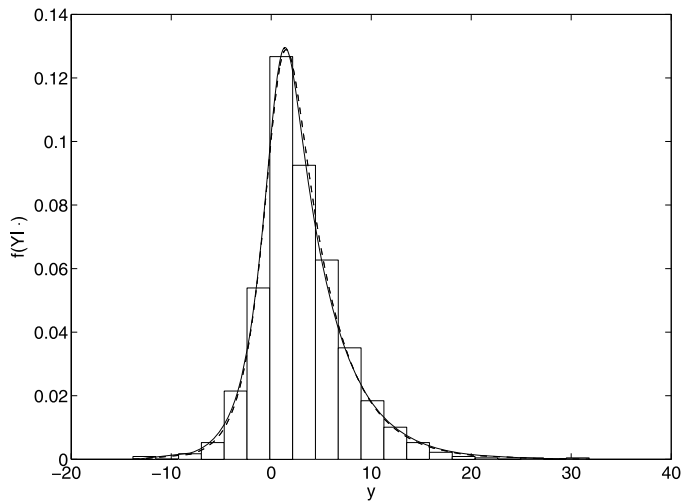}

\caption{Histogram, {fitted} (dotted line) and
theoretical (solid line) pdf for the simulated data set of Example \protect\ref{ex1}.}\label{Fig4.4}
\end{figure}

We should point out that, if instead of starting the MCMC from the
maximum likelihood estimates, we start further from this point, the
results are very similar to those obtained starting from the ML
estimates, as long as the initial value of~$\alpha$ is not very large.
It has been observed that, if the starting value of $\alpha$ is large
and the sample has long tails (small $\alpha$, as in this example),
then convergence can be extremely slow and the Gibbs algorithm often
remains stuck in the tail of the distribution for a long time. Because
of this fact we suggest starting the MCMC algorithm with small values
for $\alpha$ (not necessarily the ML estimates).

Finally, one may wonder how sensitive the method is to the
hyperparameters. We performed several analyses and our experience is
that if the real $\alpha$ or $\beta$ are not very large, then the
results are not affected by the choice of hyperparameters. For
instance, in this example, we also set $m=2$, $k=4$, $a=b=2$,
$c_{\alpha
}=c_{\beta}=0.5$, and $d_{\alpha}=d_{\beta}=0.2$, and found
$E(\bolds{\theta}
|\mathbf{ y})=(0.2609,0.5005, 1.1833,1.0157)$ with credible intervals
\begin{eqnarray*}
C_{\alpha}&=&[0.2440, 0.2927],\qquad    C_{\beta}=[0.4373, 0.5988],\\
C_{\nu}&=&[0.9188, 1.6039],\qquad   C_{\tau^2}=[0.6210, 1.6226],
\end{eqnarray*}
whose lengths are very similar to that found with the first choice of
hyperparameters. Also, the fit to the data is almost the same as in
Figure \ref{Fig4.4}. The next example illustrates the performance of
the method when $\alpha$ and/or $\beta$ are large.
\end{example}

\begin{example}\label{ex2}
Reed and Jorgensen (\citeyear{Reed}) state that if there is evidence in the
analyzed data of heavy-tailed behavior just in one tail, then it is
better to fit one of the limiting components $f_{1}$ (\ref{f1}) or
$f_{2}$ (\ref{f2}); otherwise, a frequentist approach may result in the
nonconvergence of the optimization algorithm. Here we apply the
proposed Bayesian procedure to analyze simulated data from a
$\mathit{dPlN}(\alpha,\beta,\nu,\tau^2)$ with large $\alpha$, $\beta$.
Specifically, we consider three data sets S1., S2. and S3., simulated
from $\mathit{dPlN}(10,0.5,1,1)$ (left heavy tail), $\mathit{dPlN}(0.5,10,1,1)$ (right
heavy tail) and $\mathit{dPlN}(10,10,1,1)$ (similar to a Normal distribution but
with heavier tails) distributions, respectively. We assumed the same
hyperparameters as in Example \ref{ex1}, $(m,k,a,b,c_{\alpha},c_{\beta
},d_{\alpha},d_{\beta})=(0,4,1,1,1,1,1,1)$. Table \ref
{tab:alphabetalarge} shows the starting values $\bolds{\theta}_{0}$ (ML
estimates), posterior estimates $E(\bolds{\theta}|\mathbf{ y})$ and $95\%$
credible intervals for the large parameters. We would like to point out
the high variability in the intervals, especially if $\alpha$ is large.
However, as it can be seen in Figure \ref{fig:fitlargeparameters},
both the frequentist and Bayesian approaches perform similarly when
fitting the pdf to the histogram of the data. This indicates that, as
pointed out by Reed and Jorgensen (\citeyear{Reed}), when $\alpha$ or $\beta$ are
large, the density function approaches to the three parameters limit
case $f_{2}$ (\ref{f2}) or $f_{1}$ (\ref{f1}), and, thus, there is
small difference in the $\mathit{dPlN}$ density function between multiple values
of $\alpha$ or $\beta$.

To show the versatility of the $\mathit{dPlN}$ model, we next consider two real
data sets from the insurance and internet context, respectively.
\end{example}


\begin{example}\label{ex3}
{The first data set has been analyzed in Beirlant et al. (\citeyear{Beirlant2}) and
Beirlant et al. (\citeyear{Beirlant}) and and can be found in
\href{http://lstat.kuleuven.be/Wiley/}{http://}\break
\href{http://lstat.kuleuven.be/Wiley/}{lstat.kuleuven.be/Wiley/}.
This contains 1668 claim sizes (expressed as a fraction of the sum
insured) from a fire insurance portfolio provided by the reinsurance
brokers Boels \& B\'egaul Re (AON). The data concern claim
information from office buildings. Next to the size of the claims, the
sum insured per building was provided.} The Gibbs sampler was run under
the same conditions as in the simulated-data example and posterior
estimates $E(\theta|y)=(0.51,4.99,7.78,0.76)$ were found. Note that the
posterior estimate for $\alpha$ indicates a clear long tail. Figure
\ref
{aon} shows the fit to the histogram of the data in log-scale of the
$\mathit{dPlN}$ model (solid line) in comparison with the fit provided by a
mixture of Pareto distributions (dashed line), where the number of the
components in the mixture, $k$, may change at each iteration.
%
%
\begin{table}
\caption{Starting values (MLE) and
posterior estimates for the considered simulated data \textup{S1.}, \textup{S2.} and
\textup{S3.}
in Example \textup{\protect\ref{ex2}}, where $\alpha$ or/and $\beta$ take large values. Also,
credible intervals for the large parameters are shown}\label{tab:alphabetalarge}
\begin{tabular*}{\textwidth}{@{\extracolsep{\fill}}lccc@{}}
\hline
& \multicolumn{1}{c}{\textbf{S1:} $\bolds{\mathit{dPlN}(10,0.5,1,1)}$} & \multicolumn{1}{c}{\textbf{S2:} $\bolds{\mathit{dPlN}(0.5,10,1,1)}$} &
\multicolumn{1}{c}{\textbf{S3:} $\bolds{\mathit{dPlN}(10,10,1,1)}$} \\
\hline
$\bolds{\theta}_{0}=\bolds{\theta}_{\mathit{MLE}}$ & $(4.34,0.50,
0.83,1.05)$ &
$(0.56,4.53,1.28,1.04)$ & $(4.67,5.53,0.92,0.91)$ \\
$E(\bolds{\theta}|\mathbf{ y})$ & $(22.74, 0.49,1.09,1.07)$ &
$(0.55,3.81,1.32,1.02)$ & $(40.81,1.8921,1.52,0.81)$ \\
$C_{\alpha}$ & $[1.5811,30.4547]$ & --- & $[3.0217, 50.9491]$ \\
$C_{\beta}$ & --- &$[1.7914, 9.7827]$ & $[1.4307, 3.9651]$ \\
\hline
\end{tabular*}
\end{table}
%
\begin{figure}

\includegraphics{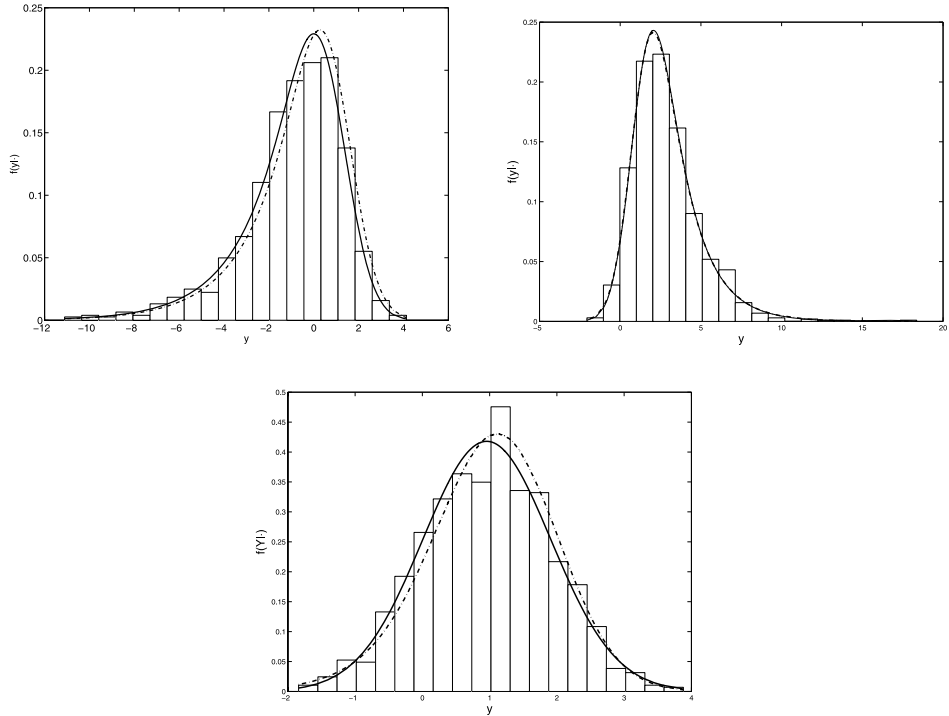}

\caption{Fitted pdfs using the ML
values (solid line) and the posterior estimates from the
Bayesian approach (dashed line), for data sets \textup{S1.}, \textup{S2.} and \textup{S3.}
in Example \textup{\protect\ref{ex2}}.} \label{fig:fitlargeparameters}
\end{figure}
Estimation for the \mbox{$k$-$\mathit{Par}$} distribution was undertaken in \citet{Ramirez},
and as it was commented in Section \ref{sec-Introduction}, here the Pareto (or
mixture of Pareto) distribution fails to capture the body of the
distribution. In addition, the Bayesian approach considered in \citet{Ramirez}
is more time consuming than the Gibbs sampler developed
here. That algorithm was based on a Birth--Death MCMC method, where at
each iteration a Metropolis--Hastings step is carried out. The Gibbs
sampler has a number of well-known advantages over standard
Metropolis--Hastings samplers. For example, the Gibbs sampler requires
no tuning, which for Metropolis--Hastings algorithms can be time
consuming---especially for long data sets where the algorithm takes
longer to run.
\end{example}

\begin{example}\label{ex4}
{The second real example that we consider is from the teletraffic
context. It can be found in the \textit{Internet Traffic Archive} (BC trace),
\href{http://www.sigcomm.org/ITA/}{http://www.sigcomm.org/ITA/},
%
\begin{figure}

\includegraphics{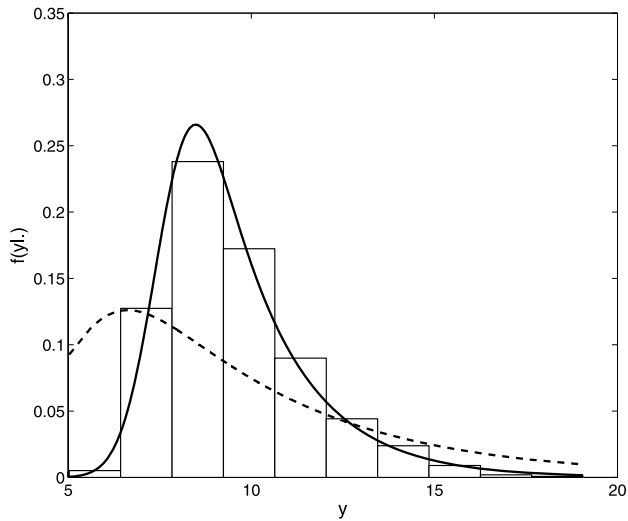}

\caption{Histogram and {fitted} pdf in Example \textup{\protect\ref{ex3}}, for the
Aon data set (claim sizes in a fire insurance portfolio) in log-scale,
under the $\mathit{dPlN}$ model (\textit{solid line}) and mixture of Pareto
components model (\textit{dashed line}).}\label{aon}
\end{figure}
where 4 million packet traces of LAN and WAN traffic seen on an
Ethernet at the Bellcore Morristown Research and Engineering facility
are recorded. The considered trace, BC-pAug89, began at 11.25 on
August~29, 1989, and ran about 3142 seconds (until 1 million packets had been
captured). The measurement techniques in making the traces are
described in Leland and Wilson (\citeyear{Leland}) and are a subset of those
analyzed in Leland et al. (\citeyear{Leland2}). The data set analyzed here consists
of the measured transferred bytes/sec within the 3142 consecutive seconds.
}

We applied the Gibbs algorithm and found posterior estimates $E(\theta
|y)=(8.59,4.52,11.83,0.59)$. The mode of this data set is not close to
zero, as can be observed in Figure \ref{t3}, and, thus, the mixture of
Pareto distributions shows a poor performance. Here again, the $\mathit{dPlN}$
model performs {well}, not only capturing the tail but also the body of
the set, as can be seen in the same figure.

Thus, from our experience the $\mathit{dPlN}$ distribution {has two advantages
over the $k$-$\mathit{Par}$} for fitting heavy-tailed data: first, it is able to
capture both the tail and body of the distribution, and second, the
estimation procedure for fitting the $\mathit{dPlN}$ distribution is faster
computationally than that proposed in \mbox{\citet{Ramirez}}, for the
$k$-$\mathit{Par}$ density.
\end{example}
%
\begin{figure}

\includegraphics{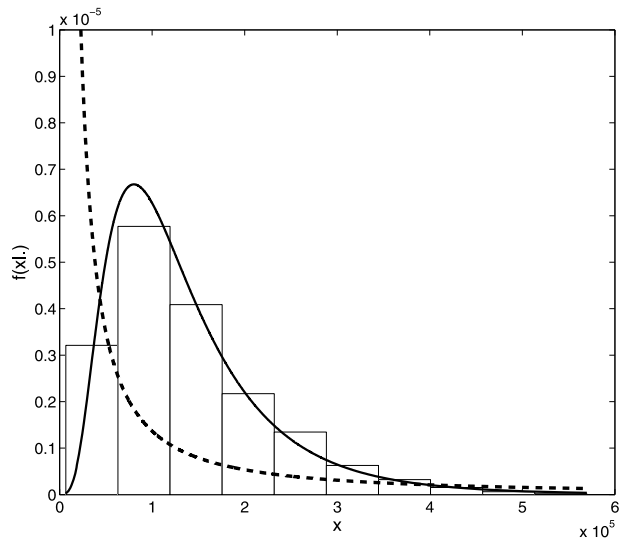}

\caption{Histogram and {fitted} pdf in Example \textup{\protect\ref{ex4}}, for the
teletraffic data set (number of bytes per second), under the $\mathit{dPlN}$
model (\textit{solid line}) and mixture of Pareto components model
(\textit{dashed line}).}\label{t3}
\end{figure}

%
\section{Inference for the $\mathit{dPlN}/M/1$ queueing system}\label{sec3}

In this section we shall consider the \textit{dPlN} distribution as
a model for the arrival process in a single-server queueing system
with independent, exponentially distributed service times. {The next
section reviews this queueing system, denoted as $\mathit{dPlN}/M/1$}.

\subsection{The $\mathit{dPlN}/M/1$ queueing system}

The $\mathit{dPlN}/M/1$ system is an example of the $G/M/1$
queueing system, whose properties are well known [see Gross and Harris
(\citeyear{Gross})]. In particular, for the $\mathit{dPlN}/M/1$ system with parameters
$\theta=(\alpha,\beta,
\nu,\tau^2)$, standard results for $G/M/1$ queues imply that the mean
interarrival time does not exist if $\alpha\leq1$. In this case, the
queueing system is automatically stable whatever the service rate $\mu$
[that is, $E(S)=1/\mu$, where $S$ denotes the service time].
Otherwise, the traffic intensity is given by
%
%
\begin{equation} \label{rhodef}\rho= \frac{(\alpha-1)(\beta
+1)}{\mu
\alpha\beta
e^{\nu+ \tau^2/2}}
.
\end{equation}

If the system is stable ($\rho<1$), then the steady-state probability
for the
number of customers $Q$ in the system just before an arrival, the
stationary time $W_{q}$ spent queueing for service, and time $W$ spent
in the system are
\begin{eqnarray*}
P(Q=n)&=&(1-r_{0})r_{0}^{n}\qquad \mbox{for all }n\in\mathbb{N} ,\\
P(W_{q}\leq x)&=&1-r_{0}e^{-\mu(1-r_{0})x},\\
P(W\leq x)&=&1-e^{-\mu(1-r_{0})x},
\end{eqnarray*}
where $r_{0}\in(0,1)$ is the unique real root of the equation
%
%
\begin{equation}
r_{0}=f^{\ast}\bigl(\mu(1-r_{0})\bigr), \label{equation_ro}
\end{equation}
and $f^{\ast}(\cdot)$ is the Laplace--Stieltjes transform of the
interarrival-time density function $f(\cdot)$ defined as
\[
f^{\ast}(s)=\int_{0}^{\infty}e^{-sx}f(x)\,dx \qquad  \mbox{for }
\operatorname{Re}(s)>0.
\]
However, the Laplace transform of the \textit{dPlN} distribution is
analytically intractable so that the standard techniques for finding
the root of Equation (\ref{equation_ro}) cannot be applied. Thus, an
alternative approach to obtaining the steady state distributions is
needed. {The next section outlines such an approach.}

\subsection{{A variety of the transform approximation method}}
The transform approximation method (\textit{TAM}) was developed informally
by Harris and Marchal (\citeyear{Harris98}) and \citet{Harris00} for the case of
approximating the Laplace transform of the single parameter Pareto
distribution and was later extended by Shortle et al. (\citeyear{Shortle04}). Here we
describe the approach in the case of the \textit{dPlN} distribution.
To approximate the Laplace transform $f^*(s)$ of the distribution of
a random variable $X$, the basic algorithm is as follows:
\begin{enumerate}
\item Pick a set of $N$ probabilities, $p_i$, $0<p_1<\cdots< p_N<1$.
\item Find the quantile $t_i$ of order $p_i$, ${P}(X\leq t_i)=p_i$.
\item Assign to each point $t_i$ the probability
\begin{eqnarray*}
w_1&=&\frac{p_1+p_2}{2},\\
w_i&=&\frac{p_{i+1}-p_{i-1}}{2}\qquad \mbox{for }i=2,\ldots,N-1,\\
w_N&=&1-\frac{p_{N-1}+p_{N}}{2}.
\end{eqnarray*}
\item{Approximate the Laplace Transform $f^*(s)$ by $f_{N}^{\ast}(s)=
\sum^{N}_{i=1}w_ie^{-st_{i}}$.}
\end{enumerate}
For the \textit{dPlN} case and once the probabilities $p_{i}$ have been
selected, the quantiles in step 2 are approximated
numerically by Newton--Raphson, with initial values obtained from the
empirical distribution function of the data.

\citet{Harris00} and Shortle et al. (\citeyear{Shortle04}) consider different
alternatives for the defining probabilities $p_i$, although, as they
point out, the choice of the optimal probabilities is an open question.
The natural
approach, known as uniform \textit{TAM} or \textit{U-TAM}, is to define uniform
probabilities, $p_i=(i-1)/N$. However, this approach leads to poor
approximations in the tail of the distribution. An alternative
algorithm applied in Shortle et al. (\citeyear{Shortle04}), which better captures
heavy-tailed behavior, is the
geometric or \textit{G-TAM} algorithm which sets $p_i = 1 - q^i$, for
$q\in(0,1)$. But even when $q\rightarrow0$, few quantiles are selected
from the body of the distribution and a poor approximation of this part
may be obtained with this approach.

We have found that a combination of both algorithms works better than
applied separately. We used the \textit{U-TAM} algorithm to obtain a
proportion $r$ of percentiles from
the body of the distribution and the \textit{G-TAM} algorithm is used to
find the other $(1-r)$ proportion of percentiles covering the
heavy tail. We consider that the body of the distribution is defined by
those percentiles $t_{i}$ such that $P(X\leq t_{i})\leq P(X\leq E[X])$,
in the case that $E[X]$ exists (otherwise, we use the median). Other
alternatives (with larger quantiles) may be used, but in practice we
have found that it makes little difference.

Formally, if $r$ denotes the proportion of percentiles before
$E[X]$, and $q$ is the geometric rate, then $(r,q)\in\{r_{\mathit{min}},\ldots
,r_{\mathit{max}}\} \times\{q_{\mathit{min}},\ldots,q_{\mathit{max}}\}$ form a grid where the
optimal value $(r^\star, q^\star)$ is chosen so that the \textit{TAM}
mean $(\sum_{i=1}^N w_{i}t_{i})$ (or the \textit{TAM} median:
$t_{a}/ \sum_{i=1}^{a}w_{i}\leq0.5$ and $\sum_{i=1}^{a+1}w_{i}> 0.5$)
matches the mean (or median) of
the original distribution. In our examples we have found that a grid of
size $8 \times17$ is enough to get a distance less than $10^{-3}$
between the \textit{TAM} mean/median and the theoretical mean/median. The
proposed methodology satisfies the conditions of Theorem
1 in Shortle et al. (\citeyear{Shortle04}) so that convergence of $f^*_N(s)$ to
$f^*(s)$ is assured as $N\rightarrow\infty$.

\subsection{{Bayesian estimation of the $\mathit{dPlN}/M/1$ queueing system}}

Given the prior distributions and a sample of
\textit{dPlN} distributed interarrival data, we have seen that the
Gibbs algorithm can be used to produce a sample of
values $\bolds{\theta}^{(t)} =
(\alpha^{(t)},\beta^{(t)},\nu^{(t)},\tau^{(t)})$ for $t=1,\ldots,T$
from the posterior distribution of the \textit{dPlN} parameters.

Supposing now that the service rate, $\mu$, is known, then it is
straightforward to estimate the probability that the system is
stable,
%
%
\begin{equation}\label{stability_probability}
P(\rho< 1|\mathbf{ y}) = \frac{1}{T} \sum_{t=1}^T
I\bigl(\rho^{(t)} < 1\bigr),
\end{equation}
where $\rho^{(t)}$ is the value of
$\rho$ calculated from Equation (\ref{rhodef}) setting $\bolds
{\theta}=\bolds{\theta}
^{(t)}$ and $I(\cdot)$ is an
indicator function. Given that this probability is high, then for
each set $\bolds{\theta}^{(t)}$ of generated parameters such that
$\rho^{(t)} < 1$, the root $r_0^{(t)}$ can be generated using
(\ref{equation_ro}) and the \textit{TAM} and, therefore, the conditional
posterior distributions of queue size and waiting times, given
stability, can be estimated by Rao Blackwellization, that is, by simply
averaging over the parameters satisfying the stability condition.
Thus, for example, the posterior distribution of queue size $
P(Q=n|\mathbf{ y})$ is estimated by
\[
\frac{1}{S} \sum_{s=1}^S P\bigl(Q=n|\bolds{\theta}^{(s)}, \mu\bigr),
\]
where $\bolds{\theta}^{(1)},\ldots, \bolds{\theta}^{(S)}$ is the
set of parameters
satisfying the stability condition.

One point to note, however, is that, as commented in Wiper (\citeyear{Wiper}), the
means of the fitted equilibrium queue size and waiting time
distributions do not exist. This is a typical
feature for Bayesian inference in $G/M/\cdot$ or $M/G/\cdot$
queueing systems. Thus, if posterior summaries of these
distributions are required, it is preferable to use the median and
quantiles.

{When the service parameter is unknown, then, given an independent
sample of service time data, conjugate inference for the service rate
can be carried out as in, for example, Armero and Bayarri (\citeyear{Armero}). For
a Monte Carlo sample, $\mu^{(1)},\ldots,\mu^{(T)}$ from the posterior
distribution of the service rate, the traffic intensity may be
estimated by calculating $\rho^{(t)}$ given $(\bolds{\theta
}^{(t)},\mu^{(t)})$
and averaging as in~(\ref{stability_probability}). In order to
condition on the existence of equilibrium, only those parameter sets
$(\bolds{\theta}^{(t)},\mu^{(t)})$ such that $\rho^{(t)}<1$ are retained.}

\subsection{Application to internet traffic analysis}
Internet traffic data has lately become a wide field of study and
numerous works have characterized it as having some unusual
statistical properties such as self similarity and heavy tails; see,
for example, Willinger, Paxson and Taqqu (\citeyear{Willinger}). In particular, as
shown in Paxson and Floyd (\citeyear{Paxson}), internet arrival traffic cannot be
well modeled by a
Poisson process. As an alternative, heavy-tailed distributions can
be considered.

\begin{figure}[b]

\includegraphics{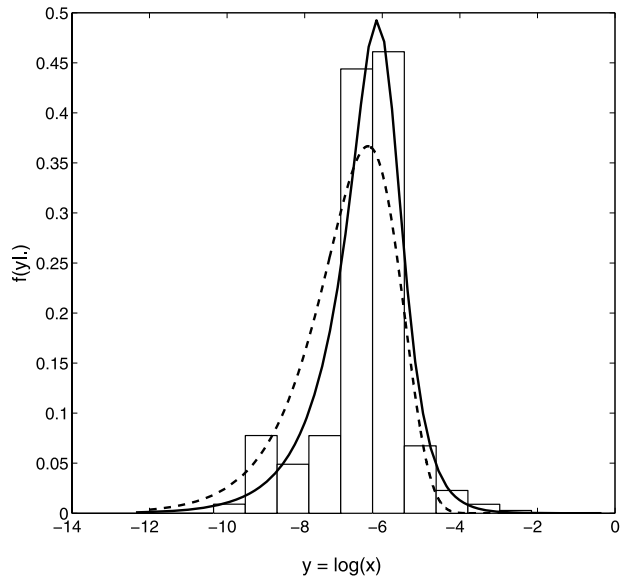}

\caption{Histograms and {fitted} pdf for the
internet data (50,000 real interarrival times) in log-scale, under the
\textit{dPlN} model (\textit{solid line}) and \textit{Pareto} model
(\textit{dashed line}).}\label{Fig4.5}
\end{figure}

{Figure \ref{Fig4.5} shows the histogram of a set of interarrival times
(in seconds) of a trace of 1 million ethernet packets, derived from
BC-pAug89 in the Internet Traffic Archive (described in Example \ref{ex3} of
Section \ref{subsec:illustration}). The first (according to the
outcome) 50,000 interarrival times (in sec) are analyzed here.
}
Superimposed (in solid line) is the {fitted} \textit{dPlN} density
generated using
the Bayesian algorithm described in Section \ref{sec2}. Also superimposed
(dashed line) is the {fitted} Pareto density. In this example the
Pareto distribution captures the tail of the distribution but has a
poorer {performance} in the body of the distribution. It can be seen in
\citet{Ramirez} that a mixture of two Pareto components provides
a good fit of this data set, however, the high computational cost of
that algorithm makes this one based on the $\mathit{dPlN}$ distribution
preferable. The posterior mean parameter
estimates for the $\mathit{dPlN}$ model were $E(\bolds{\theta}|x)=(2.15, 1.07, -6.00,
0.36)$.
%

%
\begin{table}
\caption{Probability of equilibrium and
traffic intensity. When $\mu$ is large (faster service on average), the
probability of stability of the system increases}\label{tab:internet1}
\begin{tabular*}{\textwidth}{@{\extracolsep{\fill}}ld{1.6}d{1.4}c@{}}
 \hline
\multicolumn{1}{@{}l}{$\bolds{\mu}$} & \multicolumn{1}{c}{$\bolds{E(S)}$} & \multicolumn{1}{c}{$\bolds{\mathbb{P}(\rho<1|\mathbf{ y})}$} &
\multicolumn{1}{c@{}}{$\bolds{\mathbb{E}(\rho
|\mathbf{
y})}$} \\
\hline
1500 & 0.0006 & 1 & 0.2616 \\
1000 & 0.001 & 1 & 0.3923 \\
\phantom{0}500 & 0.002 & 1 & 0.7844 \\
\phantom{0}400 & 0.0025 & 1 & 0.9798 \\
\phantom{0}395 & 0.002531 & 0.8257 & 0.9946 \\
\phantom{0}394 & 0.002538 & 0.7869 & 0.9969 \\
\phantom{0}393 & 0.002544 & 0.6115 & 0.9979 \\
\phantom{0}392 & 0.002510 & 0.4562 & 1.0008 \\
\phantom{0}391 & 0.002550 & 0.4284 & 1.0040 \\
\phantom{0}390 & 0.002564 & 0.2519 & 1.0065 \\
\phantom{0}385 & 0.002597 & 0 & 1.0194 \\ \hline
\end{tabular*}
\end{table}

\begin{figure}[b]

\includegraphics{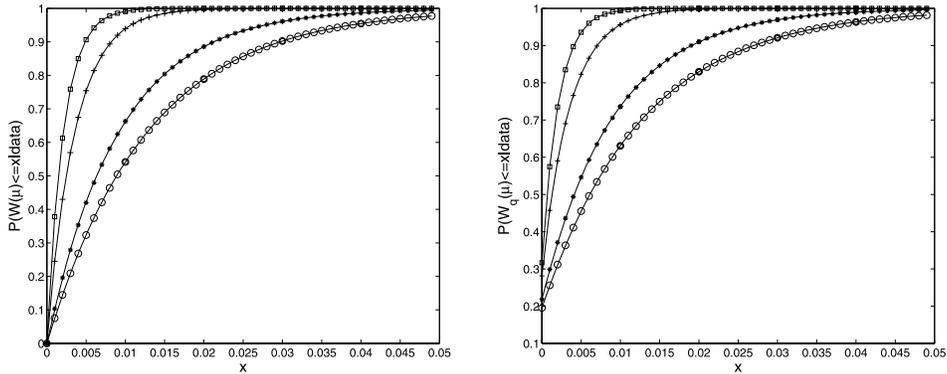}

\caption{Predictive system and queue waiting times
distributions for the internet data set for an assortment of service
rates ($\circ\mbox{: } \mu=400$, $*\mbox{: } \mu=500$, $+\mbox{: } \mu=1000$ and
$\square\mbox{: } \mu=1500$). As expected, when the service is faster ($\mu$ increases),
then the probability of waiting less than a short time is larger.}\label{Fig4.6}
\end{figure}

%
\begin{table}
\caption{Predictive system size
distribution just before an arrival for the internet data set, for an
assortment of service rates $\mu$. As expected, for faster services
(large $\mu$) the probability of an empty system is larger than for
slower services}\label{tab:internet2}
\begin{tabular*}{\textwidth}{@{\extracolsep{\fill}}lcccc@{}}
\hline
 \multicolumn{1}{@{}l}{$\bolds{\mu}$} &  \multicolumn{1}{c}{$\bolds{\mathbb{P}(Q=0)}$} &  \multicolumn{1}{c}{$\bolds{\mathbb{P}(Q=1)}$}
  &  \multicolumn{1}{c}{$\bolds{\mathbb{P}(Q=2)}$} &  \multicolumn{1}{c@{}}{$\bolds{\mathbb{P}(Q=3)}$} \\
\hline
1500 & 0.3167 & 0.2161 & 0.1475 & 0.1008 \\
1000 & 0.2813 & 0.2019 & 0.1449 & 0.1042\\
\phantom{0}500 & 0.2182 & 0.1703 & 0.1330 & 0.1039 \\
\phantom{0}400 & 0.1955 & 0.1570 & 0.1260 & 0.1014 \\
\phantom{0}395 & 0.1948 & 0.1569 & 0.1260 & 0.1013 \\
\phantom{0}394 & 0.1946 & 0.1565 & 0.1259 & 0.1013 \\
 \hline
\end{tabular*}
\end{table}

Now we shall consider the queueing aspects. Given the \textit{dPlN}
arrival process, we shall assume that arrivals are processed by a
single server with exponentially distributed service times with rate
$\mu$. Table \ref{tab:internet1} shows the posterior probability of
equilibrium (third column) and the expected value for the traffic
intensity (fourth column) for an
assortment of values of $\mu$ [the expected service time is
$E(S)=1/\mu
$]. From this table, it is clear that there is a high probability that
the system is stable (that is, no congestion occurs) for values of $\mu
$ greater than 394. Figure
\ref{Fig4.6} depicts the {fitted} system waiting time $W$, and queue
waiting time $W_{q}$, distributions for values of $\mu$ greater than
400. Table \ref{tab:internet2} illustrates the
distribution of the number $Q$ of clients in the system in equilibrium.
We can see that as the service
rate increases (i.e., the service is faster), then the median queueing
and system waiting times and the number of clients in the system
{decrease}, as would be expected.

In this example we have also compared the queueing results obtained
with the \textit{dPlN} model with those obtained from the queueing
systems $\mathit{Pareto}/M/1$ and $M/M/1$. Different estimates of the system and
queue waiting time distributions under the different queueing models
were obtained. The {fitted} system size distribution just before an
arrival among these different queues also varies, for example, the
probability that the system size is larger than 2 or than 3, $P(Q>2)$,
$P(Q>3)$ is larger with the $\mathit{dPlN}$ model than with the Pareto or
Exponential models. On the contrary, the values $P(Q>0)$, $P(Q>1)$ are
smaller with the $\mathit{dPlN}$ model than with the other ones.

\section{The $M/\mathit{dPlN}/1$ queueing system and ruin
probabilities}\label{sec4}

{In this section we consider the $M/\mathit{dPlN}/1$ queueing system, with
independent, exponentially distributed interarrival times
and \textit{dPlN} service times, and show how the Bayesian approach to
estimate the \textit{dPlN} can be used to estimate the probability of
ruin from actuarial data.}

\subsection{The $M/\mathit{dPlN}/1$ queueing system}

The general properties of the $M/G/1$
queueing system are well known; see, for example, Gross and Harris
(\citeyear{Gross}). In particular, if the service time $S$ is assumed to follow a
$\mathit{dPlN}$ distribution with $\bolds{\theta}=(\alpha,\beta,\nu,\tau
^2)$, then, if
$\alpha\leq1$, $E(S)=\infty$ and the queueing system is never
stable, whatever
the interarrival rate $\lambda$. When $\alpha> 1$, the
traffic intensity is given by
\[\label{rhomg1}
\rho=\frac{\lambda\alpha\beta e^{ \nu+
\tau^2/2}}{(\alpha-1)(\beta+1)}.
\]

The Laplace
transform $W_q^*(s)$ of the equilibrium waiting time in the queue is
related to the Laplace transform $B^*(s)$ of the (\textit{dPlN})
service time by
\[\label{mg1}
W^*_q(s)=\int_0^\infty e^{-st}\,dW_q(t)=\frac{(1-\rho)s}{s-\lambda(1-B^*(s))},
\]
where $W_q(t)$ is the distribution function of the waiting time. In
order to obtain the distribution function of the waiting time
$W_q(t)$, we first apply the \textit{TAM} to approximate $B^*(s)$ as
earlier. Second, we can use a standard numerical approach to
invert the Laplace transform, $W_q^*(s)$; see, for example, Shortle,
Fischer and Brill (\citeyear{Shortle07}) for a review. In this case, we apply the
recursion method by Fischer and Knepley (\citeyear{Fischer77}).

\begin{table}[b]
\caption{Duality between the probability of ruin in
a risk theory context\break and the $M/G/1$ queueing sytem with steady-state
queue\break waiting time distribution $W_{q}$}\label{riskqueue}
\begin{tabular*}{9cm}{@{\extracolsep{\fill}}ll@{}}
\hline
\textbf{Queueing theory} & \textbf{Risk theory} \\
\hline
Interarrival times & Interclaim times \\ [3pt]
Service times & Claim sizes \\ [3pt]
$\mathbb{P}(W_q>u)$ & Probability of ruin\\
for a $M/G/1$ &   with initial reserve $u$\\
\hline
\end{tabular*}
\end{table}

\subsection{Application to fire insurance claims}
In an insurance context, it is often assumed that claim sizes,
$C_i$, are independent and identically distributed heavy-tailed
random variables; see, for example, Rolski et al. (\citeyear{Rolski}). Here, we shall
assume that claim sizes can be modeled as \textit{dPlN} random
variables. Often, it is also supposed that the interclaim times,
$T_i$, are independent, exponentially distributed variables with
rate $\lambda$. Let $u$ denote the initial reserve of an insurance
company and let $r$ be the rate at which premium accumulates. Then,
the company's wealth, or risk portfolio at time $t,$ is
\[
R(t)=u+rt-\sum_{i=1}^{N(t)} C_i,
\]
where $N(t)= \sup(n\dvtx\sum_{i=1}^n T_i\leq t
)$ is a Poisson counting process with rate $\lambda$.

Clearly, the insurance company will be interested in the probability
that they may eventually be ruined, given their initial capital and
premium rate, that is,
%
%
\begin{equation} \label{ruinprob}
\qquad \psi(u,r) = P\bigl(R(t) < 0   \mbox{ for some $t \ge
0$}\,| \mbox{ initial capital $u$, premium rate $r$}\bigr).
\end{equation}

If the mean claim size does not exist, then eventual ruin
is certain. Otherwise, we can define the traffic intensity of this
system as $\rho= \lambda E[C_i]/r $ and it is well known that ruin
is certain if $\rho\ge1$. In the case that $\rho< 1$, then in, for
example, Prabhu (\citeyear{Prabhu}), it is shown that the ruin probability can be
computed as the steady state probability that the waiting time
exceeds $u/r$ in a $M/G/1$ queueing system, where the interarrival
time and service time distributions are the same as the
distributions of $T_i$ and $C_i/r$ respectively. Table \ref{riskqueue}
shows this duality. Thus, estimating the $M/\mathit{dPlN}/1$ queue allows us to
estimate the probability of ruin where the claims sizes are assumed to
follow a $\mathit{dPlN}$ distribution.

Note that by
scaling appropriately, it can be assumed without loss of generality
that the premium rate, $r$, is equal to $1$ and we shall do this from
now on, writing $\psi(u)$ for the ruin probability of Equation
(\ref{ruinprob}).

Assuming the $M/\mathit{dPlN}/1$ model and given some initial reserve $u$ and claim
arrival rate $\lambda$ and a sample of claim sizes, then the
posterior parameter distribution of the \textit{dPlN} claim size
distribution can be estimated using the Bayesian approach as
outlined in Section \ref{sec2} and this can be combined with the \textit{TAM} and
recursion algorithms to estimate the ruin probability.

\begin{figure}[b]

\includegraphics{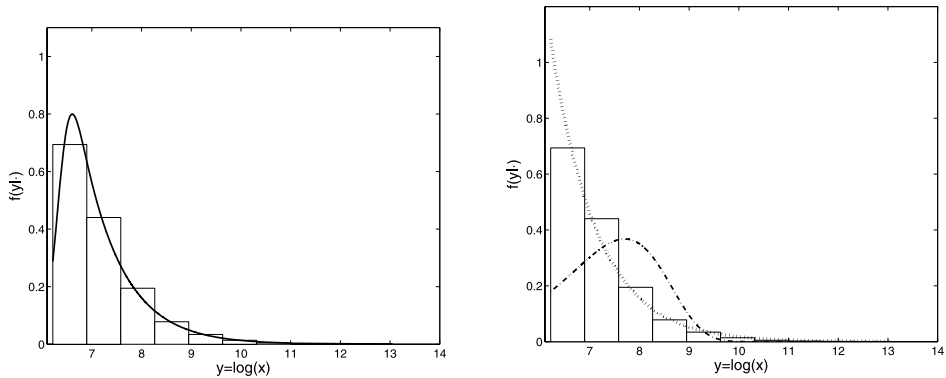}

\caption{Histograms and {fitted} pdf for the
Norwegian data (claim sizes) in log-scale, under the \textit{dPlN} (left
panel, \textit{solid line}), the Pareto (right panel, \textit{dotted line})
and Exponential (right panel, \textit{dashed line}) models.}\label{Fig4.8}
\end{figure}

To illustrate this approach, we consider data treated in Beirlant and
Goegebeur (\citeyear{Beirlant3}) and Beirlant et al. (\citeyear{Beirlant}) representing
9181 fire claims values for the period 1972--1992 from a Norwegian
insurance portfolio. Together with the year of occurrence, the values
($\times1000$ Krone) of the claims are known. They can be found in
\url{http://ucs.kuleuven.be/Wiley/index.html}.
The left panel of Figure \ref{Fig4.8} shows the data in log-scale
(values of the claims) and the Bayesian \textit{dPlN} fit. The right
panel of Figure \ref{Fig4.8} illustrates the {log-transformed} fitted
Pareto (dotted line) and Exponential (dashed line) models to this data
set. Again, the Pareto model does not capture the body of the
distribution; the Exponential fit is even worse, it captures neither
the body, nor the tail.

\begin{figure}

\includegraphics{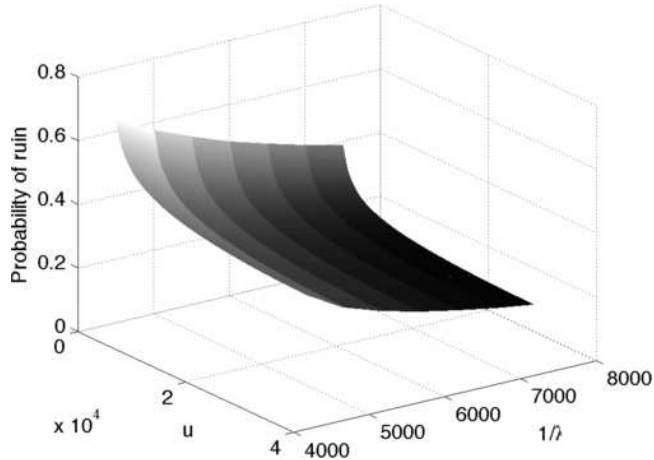}

\caption{Probabilities of ruin (z-axis) for the Norwegian
data insurance company, for an assortment of initial reserves $u$
(values from $0$ to $4\times10^{4}$) and mean interclaims times
$1/\lambda$. As would be expected, when the initial reserve is low and
the claims occur frequently on average (low values of $1/\lambda$),
then the probability of ruin increases.}\label{Fig4.9}
\end{figure}

Assuming that the system is stable, we can now estimate the
ruin probability for different interclaim rates and
initial reserves. In this case, the expected claim size,
conditional on this existing (i.e., that $\alpha> 1$), is
approximately $2915$, which implies that in order to avoid
extremely high probabilities of ruin, we should typically consider plausible
values of $\lambda$ to be below $1/2915$. Figure \ref{Fig4.9} depicts
the {posterior} probability of ruin, $E(\psi(u)|\mbox{data})$, for a
grid of values of different average interclaim times, $1/\lambda$, and
various initial reserve levels, $u$. {As would be expected, when both
the initial reserve $u$ and the expected interclaim times $1/\lambda$
are low, then the ruin probability increases.}

As we did for the $\mathit{dPlN}/M/1$ queueing system with the teletraffic data
set, given theses
claim sizes, we have also compared the performance of the $M/\mathit{dPlN}/1$
queue with the $M/\mathit{Pareto}/1$ and $M/M/1$ queueing system, assuming a
rate $\lambda=1/4000$. {When fitting a Pareto distribution to the data
with a Bayesian approach, it was found that a posteriori, the sampled
parameters of the Pareto distribution led to a lack of moment of order
one, indicating that, since $E(S|\mathbf{ y})=\infty$, then the
corresponding $M/\mathit{Pareto}/1$ system is not stable, given the data. For
the $M/M/1$ model something similar was found: $1<\rho^{(t)}<\infty$
for most of the iterations, and, thus, the posterior probability that
the system is stable was very low}. Thus, we could not predict the
probability of ruin, under these models. {Finally, the same comments as
in Section \ref{sec3}, concerning the estimation of the arrival rate $\lambda$
(interclaim times rate) when it is considered as an unknown parameter,
can be also applied here.}

\section{Conclusions}\label{sec5}
In this work we have developed Bayesian inference for the double
Pareto lognormal distribution and have illustrated that this model
can capture both the heavy-tail behavior and also the body of the
distribution for real data examples. Bayesian inference was implemented
with the Gibbs sampler, although, since $\bolds{\theta}$ is only 4
dimensional, several alternatives exist and were attempted. The use of
importance sampling was difficult because of a lack of good
distributions for the initial sample that avoided degeneracy. A block
Metropolis algorithm using a multivariate normal proposal, with
covariance matrix estimated by maximum likelihood, was also attempted
but exhibited poor mixing for $\tau$ and slower computation time. This
suggests that the Gibbs procedure should be preferred.

Second, we have combined this approach with techniques from the
queueing literature in order to estimate {posterior} equilibrium
distributions for the $\mathit{dPlN}/M/1$ and $M/\mathit{dPlN}/1$. To do this, we have
adapted the transform approximation method, in order to estimate the
Laplace transform of the \textit{dPlN} distribution and the waiting
time distribution in the $M/\mathit{dPlN}/1$ system.

Finally, we have illustrated this methodology with real data
sets, estimating first waiting times and congestion in internet
and computing the probability of ruin in the insurance context,
making use of the duality between queues and risk theory.
Comparisons with the $M/M/1$, $\mathit{Pareto}/M/1$ and $M/\mathit{Pareto}/1$ have
been also carried out. Differences among these queueing
systems, especially when the service process is heavy-tailed, were found.

A number of extensions are possible. First, we could extend our
results to the case of a multiple number of servers, that is,\ to the
$\mathit{dPlN}/M/c$ and $M/\mathit{dPlN}/c$ queueing systems or to finite capacity
systems. It would be also interesting to study the optimal control
of the systems, that is, when to open or close the queue and which
is the optimum number of servers, following the lines of Ausin, Lillo
and Wiper (\citeyear{Ausin}).

Also, in this article, we have just considered semi-Markovian
queueing systems where either the service or interarrival times
were exponential. An extension is to explore more general
distributions, in particular the so-called phase-type distributions.

{It would be interesting, too, to consider a nonparametric estimate of
the Laplace transform from data, so that a parametric specification of
the distribution entirely would be avoided. This has been suggested by
one of the referees, and will be considered in future work.}

Finally, in terms of the application to insurance, it would also be
important to explore the estimation of transient or finite time ruin
probabilities which are also of interest to insurers.

All Matlab codes and real data utilized in the examples are available
in the supplemental material Ramirez et al. (\citeyear{Ramirez2}).

\begin{appendix}
\section{\texorpdfstring{Proof of Proposition \protect\lowercase{\ref{prop1}}}{Proof of Proposition 1}}\label{appa}
For ease of notation, we write $z|y,\alpha,\beta,\nu,\tau^{2}$ as $z|y$
throughout this proof:
\begin{eqnarray*}
f_{Z|y}(z|y) & = & \frac{f_{Y,Z}(y,z)}{f_Y(y)} \\
& = & \frac{f_Z(z)f_W(y-z)}{f_Y(y)} \\
& = & \frac{1}{f_Y(y)}\frac{1}{\tau} \phi\biggl(\frac{z-\nu}{\tau} \biggr)
\frac{\alpha\beta}{\alpha+\beta} \\
&&{}\times\bigl[ \exp\bigl(\beta(y-z)\bigr)I_{z\ge
y}+\exp\bigl(-\alpha(y-z)\bigr)I_{z<y} \bigr] \\
& = &
\frac{e^{-{\nu^2}/{(2\tau^2)}}}{\tau
f_Y(y)}\frac{\alpha\beta}{\alpha+\beta}
\biggl[\exp\biggl(-\frac{1}{2\tau^2}[z^2 -2z(\nu
-\tau^2\beta)-2\tau^2\beta y ] \biggr)I_{z\ge y} \\
&&\hspace*{78pt}{}+ \exp\biggl(-\frac{1}{2\tau^2}[z^2 -2z(\nu
+\tau^2\alpha)+2\tau^2\alpha y ] \biggr)I_{z< y} \biggr]
\\
& = & \frac{e^{-{\nu^2}/{(2\tau^2)}}}{\tau
f_Y(y)}\frac{\alpha\beta}{\alpha+\beta}\\
&&{}\times \biggl[\exp\biggl(-\frac{1}{2\tau^2}\bigl[ \bigl(z-(\nu-\tau^2\beta)\bigr)^2
-2\tau^2\beta y - (\nu-\tau^2\beta)^2\bigr]\biggr)I_{z\ge y}\\
 &&\hspace*{16pt} {}+ \exp\biggl(-\frac{1}{2\tau^2}\bigl[
\bigl(z-(\nu+\tau^2\alpha)\bigr)^2 +2\tau^2\alpha y -
(\nu+\tau^2\alpha)^2\bigr]\biggr)I_{z< y}\biggr] \\
 & = &
\frac{e^{-{\nu^2}/{(2\tau^2)}}}{\tau
f_Y(y)}\frac{\alpha\beta}{\alpha+\beta}\\
&&{}\times\biggl[e^{\beta y +{(\nu-\tau^2\beta)^2}/{2\tau^2}}
\exp\biggl(-\frac{1}{2\tau^2}\bigl(z-(\nu-\tau^2\beta)\bigr)^2
\biggr)I_{z\ge
y} \\
&&\hspace*{16pt}{}+ e^{-\alpha y +{(\nu+\tau^2\alpha)^2}/{2\tau^2}}
\exp\biggl(-\frac{1}{2\tau^2}\bigl(z-(\nu+\tau^2\alpha)\bigr)^2\biggr)I_{z<
y} \biggr]
\\
& = & \frac{e^{-{\nu^2}/{(2\tau^2)}}}{
f_Y(y)}\frac{\alpha\beta}{\alpha+\beta} \biggl[ e^{\beta y +{(\nu-\tau^2\beta)^2}/{2\tau^2}}\Phi^c(y^\beta)
\frac{\phi(z^\beta)}{\tau\Phi^c(y^\beta)}I_{z\ge y} \\
&&\hspace*{78pt}{}+ e^{-\alpha
y +{(\nu+\tau^2\alpha
)^2}/{2\tau^2}}\Phi(y^\alpha)
\frac{\phi(z^\alpha)}{\tau\Phi(y^\alpha)}I_{z<
y} \biggr] \\
& = & \frac{1}{
f_Y(y)}\frac{\alpha\beta}{\alpha+\beta}
\biggl[ e^{{(2\beta y-2\nu\beta+\tau^2\beta^2)}/{2}}\Phi^c(y_{\beta})
\frac{\phi(z^\beta)}{\tau\Phi^c(y^\beta)}I_{z\ge y}\\
&&\hspace*{62pt}{} + e^{
{(-2\alpha y+2\nu\alpha+\tau^2\alpha^2)}/{2}}\Phi^c(y_{\alpha})
\frac{\phi(z^\alpha)}{\tau\Phi(y^\alpha)}I_{z<
y}\biggr] \\
& = & \frac{1}{ f_Y(y)}\frac{\alpha\beta}{\alpha+\beta}
\phi\biggl(\frac{y-\nu}{\tau} \biggr)\\
&&{}\times \biggl[
\frac{\Phi^c(y_{\beta})}{\phi(y_{\beta})}
\frac{\phi(z^\beta)}{\tau\Phi^c(y^\beta)}I_{z\ge y} +
\frac{\Phi^c(y_{\alpha})}{\phi(y_{\alpha})}
\frac{\phi(z^\alpha)}{\tau\Phi^c(y^\alpha)}I_{z< y} \biggr],
\end{eqnarray*}
which gives the
conditional density
\[
f_{Z|y}(z|y)= \biggl(R(y_{\beta})\dfrac{\phi(z^\beta)}{\tau\Phi
^c(y^\beta)} I_{z\ge y} + R(y_{\alpha})\dfrac{\phi(z^\alpha)}{\tau
\Phi
^c(y^\alpha)}I_{z<y} \biggr)\Big/{\bigl(R(y_{\alpha})+R(y_{\beta})\bigr)}.
\]
%

%
\section{\texorpdfstring{Proof of Proposition \protect\lowercase{\ref{prop2}}}{Proof of Proposition 2}}\label{appb}

Since
\[
W = E_1-E_2\qquad  \mbox{where }E_1 \sim\mathcal{E}(\alpha) \mbox{ and }
E_2 \sim\mathcal{E}(\beta),
\]
then, the distribution of $E_1|W$
is
\begin{eqnarray*}
f_{E_1|W}(e_1|w) & = &
\frac{f_{E_1,W}(e_1,w)}{f_W(w)} \\
& = &
\frac{f_{E_1,E_2}(e_1,e_1-w)}{f_W(w)} \\
& = &
\frac{f_{E_1}(e_1)f_{E_2}(e_1-w)}{f_W(w)} \\
& = &
\cases{
0, & \quad $\mbox{if }e_1 \le\max\{w,0\},$ \vspace*{2pt}\cr
\displaystyle\frac{\alpha e^{-\alpha e_1}
\beta e^{-\beta(e_1-w)}}{({\alpha\beta}/{(\alpha+\beta)})
[e^{\beta
w}I_{w< 0}+e^{-\alpha w}I_{w\ge0} ]}, & \quad $\mbox{for }e_1 >
\max\{w,0\}$}\\
& = &
\frac{(\alpha+\beta)e^{-(\alpha+\beta)e_1}}{I_{w<0}+e^{-(\alpha
+\beta
)w}I_{w\ge
0}}\qquad  \mbox{for }e_1 > \max\{w,0\}.
\end{eqnarray*}
%

\section{\texorpdfstring{Proof of Proposition \protect\lowercase{\ref{prop3}}}{Proof of Proposition 3}}\label{appc}
Note first that
\begin{eqnarray*}
P(w_{1}>0,\ldots,w_{n}>0|\mathbf{ y},\bolds{\theta}) &=&
P(z_{1}<y_{1},\ldots
,z_{n}<y_{n}|\mathbf{ y},\bolds{\theta})
\\
&=& \prod_{i=1}^n \Phi
\biggl(\frac{y_{i}-\nu
}{\tau} \biggr) > 0
\end{eqnarray*}
for any set $\mathbf{ y}$ and where $\Phi$ is the standard normal
cumulative distribution. Therefore,
\begin{eqnarray*}
P(w_{1}>0,\ldots,w_{n}>0|\mathbf{ y}) &=& \int P(w_{1}>0,\ldots
,w_{n}>0|\mathbf{
y},\bolds{\theta}) f(\bolds{\theta}|\mathbf{ y})   \,d\bolds{\theta}>0
\end{eqnarray*}
for any $\mathbf{ y}$. Similarly, $P(w_{1}<0,\ldots,w_{n}<0|\mathbf{ y}) > 0$
for any $\mathbf{ y}$.

Now consider the posterior distribution of $\alpha,\beta|\mathbf{ w}$,
\begin{eqnarray*}
f(\alpha,\beta|\mathbf{ w}) & \propto& f(\mathbf{
w}|\alpha
,\beta)f(\alpha,\beta) \\
& \propto& \biggl( \frac{\alpha\beta
}{\alpha
+\beta}\biggr)^n \exp\Biggl(\beta\sum_{i=1}^n
w_{i}I(w_{i}<0)\Biggr)\exp
\Biggl(-\alpha\sum_{i=1}^n w_{i}I(w_{i}>0)\Biggr)\\
&&{}\times f(\alpha,\beta).
\end{eqnarray*}
In the case that all $w_{i}<0$, then when $\alpha\rightarrow\infty$,
for any given $\beta$,
\[
f(\alpha|\beta,\mathbf{ w}) \propto f(\alpha,\beta|\mathbf{ w}) \rightarrow
c(\beta) f(\alpha|\beta)
\]
for some $c(\beta)>0$. Equally, if all $w_{i}>0$, then when $\beta
\rightarrow\infty$, for any given $\alpha$,
\[
f(\beta|\alpha,\mathbf{ w}) \propto f(\alpha,\beta|\mathbf{ w}) \rightarrow
d(\alpha) f(\beta|\alpha)
\]
for some $d(\alpha)>0$.
Therefore, if
$\int_{a}^{\infty} f(\alpha|\beta) \,d\alpha$ is divergent for any
$a \ge
0$, then we have immediately that
when $\alpha\rightarrow\infty$, $f(\alpha|\mathbf{ w},\beta)
\rightarrow
c(\beta) f(\alpha|\beta)$, which implies that the posterior
distribution of $\alpha$ is improper and similarly in the case of an
improper prior for $\beta|\alpha$.
\end{appendix}

\section*{Acknowledgments}
The authors are grateful to three anonymous reviewers for their
detailed and insightful comments on an earlier version.

\begin{supplement}[id=suppA]
\sname{Supplement}
\stitle{Matlab Toolbox}
\slink[doi]{10.1214/10-AOAS336SUPP}
\slink[url]{http://lib.stat.cmu.edu/aoas/336/supplement.zip}
\sdatatype{.zip}
\sdescription{The Matlab toolbox performs Bayesian estimation for the
double Pareto Lognormal ($\mathit{dPlN}$) distribution, and for the queueing
systems $\mathit{dPlN}/G/1$ and $M/\mathit{dPlN}/1$.}
\end{supplement}

\printaddresses


\begin{thebibliography}{9}

\bibitem[\protect\citeauthoryear{Adler, Feldman and Taqqu}{1999}]{Adler}
\textsc{Adler, R.}, \textsc{Feldman, R.} and \textsc{Taqqu, M. T.}
(1999). \textit{A Practical Guide to Heavy Tails: Statistical
Techniques and Applications}. Birkh\"{a}user, Boston.
\MR{1652283}

\bibitem[\protect\citeauthoryear{}{1994}]{Armero}
\textsc{Armero, C.} and \textsc{Bayarri, M. J.} (1994).
Bayesian prediction in $M/M/1$ queues. \textit{Queueing Syst.}
\textbf{15} 401--417.
\MR{1266803}

\bibitem[\protect\citeauthoryear{Aus\'{\i}n, Lillo and Wiper}{2007}]{Ausin}
\textsc{Aus\'{\i}n, M. C.}, \textsc{Lillo, R. E.} and \textsc{Wiper,
M. P.} (2007).
Bayesian control of the number of
servers in a $GI/M/c$ queueing system. \textit{J. Statist.
Plann. Inference}
\textbf{137} 3043--3057.
\MR{2364149}

\bibitem[\protect\citeauthoryear{}{1998}]{Beirlant2}
\textsc{Beirlant, J.}, \textsc{Goegebeur, Y.}, \textsc{Verlaak, R.} and
\textsc{Vynckier, P.} (1998).
Burr regression and portfolio segmentation. \textit{Insurance
Math. Econom.}
\textbf{23} 231--250.

\bibitem[\protect\citeauthoryear{}{2003}]{Beirlant3}
\textsc{Beirlant, J.} and \textsc{Goegebeur, Y.} (2003).
Regression with response distributions of Pareto-type. \textit
{Computat. Statist. Data Anal.}
\textbf{42} 595--619.
\MR{1967059}

\bibitem[\protect\citeauthoryear{}{2004}]{Beirlant}
\textsc{Beirlant, J.}, \textsc{Goegebeur, Y.}, \textsc{Segers, J.} and
\textsc{Teugels, J.} (2004).
\textit{Statistics of Extremes: Theory and Applications.}
Wiley, New York.
\MR{2108013}

\bibitem[\protect\citeauthoryear{}{1973}]{boxtiao}
\textsc{Box, G.} and \textsc{Tiao, G.} (1973).
\textit{Bayesian Inference in Statistical Analysis.}
Wiley, New York.

\bibitem[\protect\citeauthoryear{Embrechts, Kl\"{u}ppelberg and Mikosch}{1997}]{Embrechts}
\textsc{Embrechts, P.}, \textsc{Kl\"{u}ppelberg, C.} and \textsc
{Mikosch, T.} (1997). \textit{Modelling Extremal Events for Insurance and
Finance}.
Springer, Heidelberg.
\MR{1458613}

\bibitem[\protect\citeauthoryear{}{1977}]{Fischer77}
\textsc{Fischer, M.} and \textsc{Knepley, J.} (1977).
A numerical solution for some computational problems occurring in
queueing theory. In \textit{Algorithmic Methods in Probability, Studies in
Management
Science} 271--285. North-Holland, Amsterdam.

\bibitem[\protect\citeauthoryear{}{1998}]{Gross}
\textsc{Gross, D.} and \textsc{Harris, C. M.} (1998).
\textit{Fundamentals of Queueing
Theory}. Wiley, New York.
\MR{1600527}

\bibitem[\protect\citeauthoryear{}{1998}]{Harris98}
\textsc{Harris, C. M.} and \textsc{Marchal, W. G.} (1998).
Distribution estimation using Laplace transforms. \textit{INFORMS J.
Comput.} \textbf{10} 448--458.
\MR{1656928}

\bibitem[\protect\citeauthoryear{Harris, Brill and Fischer}{2000}]{Harris00}
\textsc{Harris, C. M.}, \textsc{Brill, P. H.} and \textsc{Fischer, M. J.}
(2000). Internet-type queues with power-tailed interarrival times and
computational
methods for their analysis. \textit{INFORMS J. Comput.} \textbf
{12} 261--271.

\bibitem[\protect\citeauthoryear{}{1991}]{Leland}
\textsc{Leland, W. E.} and \textsc{Wilson, D. V.} (1991). High
time-resolution measurement and analysis of LAN traffic: Implications
for LAN interconnection. In \textit{Proc. IEEE INFOCOM'91} 1360--1366.
Bat Harbour, FL.

\bibitem[\protect\citeauthoryear{}{1994}]{Leland2}
\textsc{Leland, W. E.}, \textsc{Taqqu, M.}, \textsc{Willinger, W.} and
\textsc{Wilson, D. V.} (1994). On the self-similar nature of Ethernet
traffic (extended version). \textit{IEEE/ACM Transactions on
Networking} \textbf{2} 1--15.

\bibitem[\protect\citeauthoryear{}{1995}]{Paxson}
\textsc{Paxson, V.} and \textsc{Floyd, S.} (1995).
Wide area traffic: The failure of Poisson modeling. \textit{IEEE/ACM
Transactions on Networking} \textbf{3} 236--244.

\bibitem[\protect\citeauthoryear{}{1985}]{Philbrick}
\textsc{Philbrick, S. W.} (1985).
A practical guide to the single parameter
Pareto distribution. In \textit{Proceedings of the Casualty Actuarial Society}
\textbf{LXXII} 44--123. Boca Raton, FL.

\bibitem[\protect\citeauthoryear{}{1998}]{Prabhu}
\textsc{Prabhu, N. U.} (1998). \textit{Stochastic Storage Processes: Queues,
Insurance Risk, Dams, and Data Communication}. Springer, Berlin.
\MR{1492990}

\bibitem[\protect\citeauthoryear{Ram\'irez, Lillo and Wiper}{2008}]{Ramirez}
\textsc{Ram\'irez, P.}, \textsc{Lillo, R. E.} and \textsc{Wiper,
M. P.} (2008).
Bayesian analysis of a queueing system with a long-tailed arrival
process. \textit{Comm. Statist. Simulation Comput.} \textbf{4} 697--712.

\bibitem[\protect\citeauthoryear{}{2010}]{Ramirez2}
\textsc{Ram\'irez, P.}, \textsc{Lillo, R. E.}, \textsc{Wilson, S.} and
\textsc{Wiper, M. P.} (2010).
Supplement to ``Bayesian inference for double Pareto Lognormal
queues.'' DOI: \href{http://dx.doi.org/10.1214/10-AOAS336SUPP}{10.1214/10-AOAS336SUPP}.

\bibitem[\protect\citeauthoryear{}{2004}]{Reed}
\textsc{Reed, W. J.} and \textsc{Jorgensen, M.} (2004).
The double Pareto--lognormal distribution---A new parametric model for
size distributions. \textit{Comm. Statist. Theory Methods} \textbf{33} 1733--1753.
\MR{2065171}

\bibitem[\protect\citeauthoryear{}{1997}]{Resnick}
\textsc{Resnick, S. I.} (1997).
Heavy tail modeling and teletraffic data. \textit{Ann.
Statist.} \textbf{25} 1805--1848.
\MR{1474072}

\bibitem[\protect\citeauthoryear{}{2001}]{Robert}
\textsc{Robert, C. P.} (2001). \textit{The Bayesian Choice}. Springer, New York.
\MR{1835885}

\bibitem[\protect\citeauthoryear{}{1999}]{Rolski}
\textsc{Rolski, T.}, \textsc{Schmidli, H.}, \textsc{Schmidt, V.} and
\textsc{Teugels, J.} (1999). \textit{Stochastic Processes for Insurance
and Finance}. Wiley, Chichester.
\MR{1680267}

\bibitem[\protect\citeauthoryear{}{2004}]{Shortle04}
\textsc{Shortle, J. F.}, \textsc{Brill, P. H.}, \textsc{Fischer, M.
J.},
\textsc{Gross, D.} and \textsc{Massi, D. M. B.} (2004).
An algorithm to compute the waiting time distribution
for the $M/G/1$ queue. \textit{INFORMS J. Comput.} \textbf{16} 52--161.
\MR{2065995}

\bibitem[\protect\citeauthoryear{Shortle, Fischer and Brill}{2007}]{Shortle07}
\textsc{Shortle, J. F.}, \textsc{Fischer, M. J.} and \textsc{Brill,
P. H.} (2007).
Waiting-time distribution of $M/D_N/1$ queues through numerical Laplace
inversion. \textit{INFORMS J. Comput.} \textbf{19} 112--120.
\MR{2300590}

\bibitem[\protect\citeauthoryear{}{1999}]{Sigman}
\textsc{Sigman, K.} (1999).
A primer on heavy-tailed distributions. \textit{Queueing Syst.}
\textbf{33} 261--275.
\MR{1748646}

\bibitem[\protect\citeauthoryear{Willinger, Paxson and Taqqu}{1998}]{Willinger}
\textsc{Willinger, W.}, \textsc{Paxson, V.} and \textsc{Taqqu, M. S.} (1998).
Self-similarity and heavy tails: Structural modeling of network
traffic. In \textit{A Practical Guide to Heavy Tails: Statistical
Techniques and
Applications} (R. Adler, R. Feldman and M. S. Taqqu, eds.) 27--54.
Birkh\"{a}user, Boston.
\MR{1652283}

\bibitem[\protect\citeauthoryear{}{1997}]{Wiper}
\textsc{Wiper, M. P.} (1997).
Bayesian analysis of $Er/M/1$ and $Er/M/c$ queues. \textit{J.
Statist. Plann. Inference} \textbf{69} 65--79.
\MR{1631145}

\end{thebibliography}
\end{document}